\documentclass[aps,preprint,showpacs,preprintnumbers,amsmath,amssymb]{revtex4}

\usepackage{dcolumn}
\usepackage{mathrsfs}
\usepackage{bm}
\usepackage{amsmath,amssymb,epsfig,float}

\begin{document}

\title{Entanglement redistribution in the Schwarzschild spacetime}

\author{Jieci Wang, Qiyuan Pan and Jiliang Jing\footnote{Corresponding author, Email: jljing@hunnu.edu.cn}}
\affiliation{ Institute of Physics and  Department of Physics,
\\ Hunan Normal University, Changsha, \\ Hunan 410081, P. R. China
\\ and
\\ Key Laboratory of Low-dimensional Quantum Structures
\\ and Quantum
Control of Ministry of Education, \\ Hunan Normal University,
Changsha, Hunan 410081, P. R. China}

\vspace*{0.2cm}
\begin{abstract}
\vspace*{0.2cm} The effect of Hawking radiation on the
redistribution of the entanglement and mutual information in the
Schwarzschild spacetime is investigated. Our analysis shows that the
physically accessible correlations degrade while the unaccessible
correlations increase as the Hawking temperature increases because
the initial correlations described by inertial observers are
redistributed between all the bipartite modes. It is interesting to
note that, in the limit case that the temperature tends to infinity, the
accessible mutual information equals to just half of its initial
value, and the unaccessible mutual information between mode $A$ and
$II$ also equals to the same value.

\end{abstract}

\vspace*{1.5cm}

\maketitle

\section{introduction}

Entanglement plays a pivotal role in quantum information | it is a
resource for various computational tasks such as quantum
communication and teleportation. It is believed that investigation of
the entanglement in a relativistic framework is not only helpful in
understanding some of the key questions in quantum information
theory, but also plays an important role in the study of entropy and
the information paradox of black holes \cite{Ahn,Bombelli-Callen,
Hawking-Terashima}. Thus, much attention has been focused on the
relativistic effects in the context of the quantum information
theory \cite{Peres,Boschi,Bouwmeester,
Alsing-Milburn,Alsing-McMahon-Milburn,
Ge-Kim,Schuller-Mann,Alsing-Mann, moradi,jieci,
adesso,Pan,jieci1,jieci2,David}. Recently, we investigated the
effect of the Hawking radiation \cite{Hawking-1} on the entanglement
and teleportation in a general static and asymptotically flat black
hole with spherical symmetry \cite{Pan Qiyuan}, and found that the
entanglement degraded as the increase of the Hawking temperature
both for the scalar and Dirac fields. However, we now face a
intriguing question: where has the lost entanglement gone?

In this paper we will investigate the redistribution of entanglement  for
the Dirac fields in the background of a Schwarzschild black hole. The
loss of entanglement will be explained by the redistribution of the
entanglement among all accessible and unaccessible modes. We use
some methods of the quantum information to quantify and identify the
property of the correlations (both quantum and classical) from the
perspective of physical observers who can access field modes only
outside the event horizon. Our scheme proposes that two observers,
Alice and Bob, share a generically entangled state at the same
initial point in flat region. After their coincidence, Alice remains
at the asymptotically flat region but Bob freely falls in toward the
black hole and locates near the event horizon. Due to the presence
of a horizon, an observer in each side of the horizon has no access
to field modes in the causally disconnected region. Therefore, the
observer must trace over the inaccessible region and lose some
information about the state. Thus we must calculate the entanglement
in all possible bipartite divisions of the system: (i) the mode $A$
described by Alice, (ii) the mode $I$ in exterior region of the
black hole (described by Bob), and (iii) the complimentary mode $II$
in the interior region of the black hole.

The outline of the paper is as follows. In Sec. II we recall the vacuum structure and Hawking radiation for the Dirac
fields. In Sec. III we discuss the essential features of the
background spacetime and the redistribution of entanglement. We will
summarize and discuss our conclusions in the last section.

\section{Vacuum structure and Hawking Radiation for Dirac
fields} The line element for the Schwarzschild spacetime is
\begin{eqnarray}\label{matric}
ds^2=-(1-\frac{2M}{r}) dt^2+(1-\frac{2M}{r})^{-1} dr^2+r^2(d\theta^2
+sin^2\theta d\varphi^2),
\end{eqnarray}
where the parameter $M$ represents the mass of the black hole.

Solving the Dirac equation \cite{Brill} near the event horizon, we
obtain the positive (fermions) frequency outgoing solutions outside
and inside regions of the event horizon $r_+$ \cite{jing1,jing2}
\begin{eqnarray}\label{outside mode}
\Psi_{\mathbf{k}}(r>r_{+})=\mathcal {G}e^{-i\omega u},
\end{eqnarray}
\begin{eqnarray}\label{inside mode}
\Psi_{\mathbf{k}}(r<r_{+})=\mathcal {G}e^{i\omega u},
\end{eqnarray}
where $\mathcal{G}$ is a 4-component Dirac spinor \cite{jieci1},
$u=t-r_{*}$ and $r_{*}=r+2M\ln\frac{r-2M}{2M}$ is the tortoise coordinate.
Hereafter we will use the wavevector $\mathbf{k}$ labels the modes.
Particles and antiparticles will be classified with respect to the
future-directed timelike Killing vector in each region.

Making a analytic continuation for Eqs. (\ref{outside mode}) and
(\ref{inside mode}), we find a complete basis for positive energy
modes according to Domour-Ruffini's suggestion  \cite{D-R}. Then
we can quantize the Dirac fields in the Schwarzschild  and Kruskal
modes respectively, from which we can easily get the Bogoliubov
transformations \cite{Barnett} between the creation and annihilation
operators in the Schwarzschild and Kruskal coordinates. After
properly normalizing the state vector, the vacuum state of the
Kruskal particle  for mode $\mathbf{k}$ is found to be
\begin{eqnarray}\label{Dirac-vacuum}
\nonumber|0\rangle_{K}&=&
(e^{-\omega_\mathbf{k}/T}+1)^{-\frac{1}{2}}\exp\bigg[e^{-\omega_\mathbf{k}/2T}a^{I\dag}_{\mathbf{k}}b^{II\dag}
_{\mathbf{-k}}\bigg]|0_{\mathbf{k}}\rangle^{+}_{I}|0_{-\mathbf{k}}\rangle
^{-}_{II}\\
\nonumber&=&\bigg[(e^{-\omega_\mathbf{k}/T}+1)^{-\frac{1}{2}}+(e^{\omega_\mathbf{k}/T}+1)^{-\frac{1}{2}}
a^{I\dag}_{\mathbf{k}}b^{II\dag}_{\mathbf{-k}}\bigg]|0_{\mathbf{k}}\rangle^{+}_{I}|0_{-\mathbf{k}}\rangle
^{-}_{II}\\
&=&(e^{-\omega_\mathbf{k}/T}+1)^{-\frac{1}{2}}|0_{\mathbf{k}}\rangle^{+}_{I}|0_{-\mathbf{k}}\rangle
^{-}_{II}+(e^{\omega_\mathbf{k}/T}+1)^{-\frac{1}{2}}|1_{\mathbf{k}}\rangle^{+}_{I}|1_{-\mathbf{k}}\rangle
^{-}_{II}.
\end{eqnarray}
where $\{|n_{-\mathbf{k}}\rangle^{-}_{II}\}$ and
$\{|n_{\mathbf{k}}\rangle^{+}_{I}\}$ are the orthonormal bases for
the inside and outside regions of the event horizon respectively,
the superscript  $\{+,-\}$ on the kets is used to indicate the
particle and antiparticle vacua, and $T=\frac{1}{8\pi M}$ is the
Hawking temperature \cite{Kerner-Mann}. The only excited state is
\begin{eqnarray}\label{Dirac-excited}
&&|1\rangle_{K}=
|1_{\mathbf{k}}\rangle^{+}_{I}|0_{-\mathbf{k}}\rangle^{-}_{II}.
\end{eqnarray}
Hereafter we will refer to the  particle mode
$\{|n_{\mathbf{k}}\rangle^{+}_{I}\}$ simply as $\{|n\rangle_{I}\}$,
and the antiparticle mode $\{|n_{-\mathbf{k}}\rangle^{-}_{II}\}$ as
$\{|n\rangle_{II}\}$.

\section{Entanglement redistribution}

We assume that Alice has a detector which only detects mode
$|n\rangle_{A}$ and Bob has a detector sensitive only to mode
$|n\rangle_{B}$, and they share a generically entangled state at the
same initial point in flat Minkowski spacetime. The generically
entangled initial state is
\begin{eqnarray}\label{initial}
|\Psi\rangle_{AB}=\alpha|0\rangle_{A}|0\rangle_{B}
+\sqrt{1-\alpha^{2}}|1\rangle_{A}|1\rangle_{B},
\end{eqnarray}
where $\alpha$ is a state
parameter which satisfies
$|\alpha|\in(0,1)$. After the coincidence of Alice and Bob, Alice stays
stationary at the asymptotically flat region, while Bob falls toward
the black hole and hovers outside the event horizon. Using Eqs.
(\ref{Dirac-vacuum}) and (\ref{Dirac-excited}), we can rewrite Eq.
(\ref{initial}) in terms of Minkowski modes for Alice and black hole
modes for Bob
\begin{eqnarray}  \label{state}
\nonumber|\Psi\rangle_{A,I,II}&=&\alpha|0\rangle_{A}\bigg[(e^{-\omega_\mathbf{k}/T}+2)^{-\frac{1}{2}}
|0\rangle_{I}|0\rangle_{II}+(e^{\omega_\mathbf{k}/T}+2)^{-\frac{1}{2}}
|1\rangle_{I}|1\rangle_{II}\bigg]\\&&+\sqrt{1-\alpha^2}|1\rangle_{A}
|1\rangle_{I}|0\rangle_{II}.
\end{eqnarray}

\subsection{The physically accessible correlations}

Since the exterior region is causally disconnected from the interior
region of the black hole, the only entanglement which is physically
accessible to the observers is encoded in the mode $A$ described by
Alice and the mode $I$ in exterior region of the black hole
described by Bob. Thus, when  observers describe the state they find
that some of the correlations are lost \cite{Pan Qiyuan}. Taking the
trace over the state of the interior region we obtain
\begin{eqnarray}  \label{eq:state1}
\nonumber\varrho_{A,I}=&&\alpha^2(e^{-\omega_\mathbf{k}/T}+1)^{-1}|00\rangle\langle00|+\alpha\sqrt{1-\alpha^2}
(e^{-\omega_\mathbf{k}/T}+1)^{-\frac{1}{2}}(|00\rangle\langle11|+|11\rangle\langle00|)\\
&&+\alpha^2(e^{\omega_\mathbf{k}/T}+1)^{-1}|01\rangle\langle01|+(1-\alpha^2)|11\rangle\langle11|
\end{eqnarray}
where $|mn\rangle=|m\rangle_{A}|n\rangle_{B,I}$. The partial
transpose criterion provides a necessary and sufficient condition
for the entanglement in a mixed state of two qubits \cite{Peres}: if
at least one eigenvalue of the partial transpose is negative, the
density matrix is entangled. The partial transpose
$\varrho_{AB}^{T_{A}}$ is obtained by interchanging Alice's qubits,
which yields a negative eigenvalue
\begin{eqnarray*}
\lambda_{-}=\frac{1}{2}\left[\alpha^{2}(e^{\omega/T}+1)^{-1}
-\sqrt{\alpha^{4}(e^{\omega/T}+1)^{-2}+4\alpha^{2}(1-\alpha^{2})(e^{-\omega/T}+1)^{-1}}\right].
\end{eqnarray*}
Thus, the state is always entangled for any Hawking temperature $T$.
To quantify the entanglement of $\rho_{A,I}$ in
Eq.~(\ref{eq:state1}) we compute the spin-flip matrix
$\tilde{\varrho}_{A,I}$, and find that the eigenvalues of the matrix
$\varrho_{A,I}\tilde{\varrho_{A,I}}$ are
$\bigg[4\alpha^2(1-\alpha^2)(e^{-\omega_\mathbf{k}/T}+1)^{-1},0,0,0\bigg]$.
Then we find the concurrence \cite{Wootters,Coffman} of this state
\begin{eqnarray}
C(\varrho_{A,I})=2\alpha\sqrt{1-\alpha^2}(e^{-\omega_\mathbf{k}/T}+1)^{-\frac{1}{2}},
\end{eqnarray} which
is $2\alpha\sqrt{1-\alpha^2}$ at zero Hawking temperature, i.e., the
case of supermasive or an almost extreme black hole, as expected.
And approaches the value $C_f(\varrho
_{A,I})=\alpha\sqrt{2(1-\alpha^2)}$ for infinite Hawking temperature
$T\rightarrow \infty$,  i.e., the black hole evaporates completely.
The entanglement of formation \cite{Wootters} is
\begin{eqnarray*}
E_{F}(\varrho_{A,I})=\mathcal
{H}\bigg[\frac{1+\sqrt{1-4\alpha^2(1-\alpha^2)(e^{-\omega_\mathbf{k}/T}+1)^{-1}}}{2}\bigg],
\end{eqnarray*}
where $\mathcal{H}(x)=-x \log_2 x -(1-x)\log_2(1-x)$.

The mutual information \cite{RAM}, which can be used to estimate the
total (classical and quantum) amount of correlations between any two
subsystems of the overall system, is found to be
\begin{eqnarray}
I(\varrho_{A,I})&=&\mathcal
{F}[1-\alpha^{2}(e^{\omega/T}+1)^{-1}]+\mathcal
{F}[\alpha^{2}(e^{\omega/T}+1)^{-1}]-\mathcal{F}(1-\alpha^{2})
\nonumber\\&&-\mathcal
{F}[1-\alpha^{2}(e^{-\omega/T}+1)^{-1}]-\mathcal{F}[\alpha^{2}(e^{-\omega/T}+1)^{-1}]-\mathcal
{F}(\alpha^{2}),
\end{eqnarray} where $\mathcal{F}(x)=x\log(x)$. Note that the
initial mutual information is
$I_{i}(\varrho_{A,I})=-2[\mathcal{F}(\alpha^{2})
+\mathcal{F}(1-\alpha^{2})]$ for vanishing Hawking temperature. In
the infinite Hawking temperature limit $T\rightarrow\infty$, the
mutual information converges to
$I_{f}(\varrho_{A,I})=-[\mathcal{F}(\alpha^{2})
+\mathcal{F}(1-\alpha^{2})]$, which is just half of $I_{i}$.

\subsection{The physically unaccessible correlations}

To explore entanglement in this system in more detail we consider
the tripartite system consisting of the modes $A$, $I$, and $II$. In
an inertial frame the system is bipartite, but from a non-inertial
perspective an extra set of modes in region $II$ becomes relevant.
We therefore calculate the entanglement in all possible bipartite
divisions of the system. Let us first comment on the quantum
correlations created between the mode $A$ and mode $II$, tracing
over the mode in region $I$, we obtain the density matrix
\begin{eqnarray}  \label{eq:state}
\nonumber\varrho_{A,II}&=&\alpha^2(e^{-\omega_\mathbf{k}/T}+1)^{-1}|00\rangle\langle00|+\alpha\sqrt{1-\alpha^2}
(e^{\omega_\mathbf{k}/T}+1)^{-\frac{1}{2}}(|10\rangle\langle01|+|01\rangle\langle10|)\\
&&+\alpha^2(e^{\omega_\mathbf{k}/T}+1)^{-1}|01\rangle\langle01|
+(1-\alpha^2)|10\rangle\langle10|,
\end{eqnarray}
where $|mn\rangle=|m\rangle_{A}|n\rangle_{B,II}$. The partial
transpose of $\varrho_{A,II}$  has an eigenvalue \begin{eqnarray*}
\lambda_{-}=\frac{1}{2}\left[\alpha^{2}(e^{-\omega/T}+1)^{-1}
-\sqrt{\alpha^{4}(e^{-\omega/T}+1)^{-2}+4\alpha^{2}(1-\alpha^{2})(e^{\omega/T}+1)^{-1}}\right],
\end{eqnarray*} which is less than or equal to zero.  At $T =0$ the
eigenvalue is zero, which means that there is no entanglement at
this point. However, for $T>0$ entanglement does exist between these
two modes.

Calculating the spin-flip of $\varrho_{A,II}$
\begin{eqnarray}
\nonumber\tilde{\varrho}_{A,II}=&&\alpha^2(e^{-\omega_\mathbf{k}/T}+1)^{-1}|11\rangle\langle11|+\alpha\sqrt{1-\alpha^2}
(e^{\omega_\mathbf{k}/T}+1)^{-\frac{1}{2}}(|01\rangle\langle10|+|10\rangle\langle01|)\\&&+\alpha^2
(e^{\omega_\mathbf{k}/T}+1)^{-1}|10\rangle\langle10|+(1-\alpha^2)
|01\rangle\langle01|,
\end{eqnarray}%
we find that
\begin{eqnarray}
\nonumber\varrho_{A,II}\,\tilde{\rho}_{A,II}=2\alpha^2(1-\alpha^2)(e^{\omega_\mathbf{k}/T}+1)^{-1}
(|01\rangle\langle01|+|10\rangle\langle10|)+2\sqrt{\alpha^6(1-\alpha^2)}\\
\times(e^{\omega_\mathbf{k}/T}+1)^{-\frac{3}{2}}|01\rangle\langle10|+2\sqrt{\alpha^2(1-\alpha^2)^3}
(e^{\omega_\mathbf{k}/T}+1)^{-\frac{1}{2}}|10\rangle\langle01|,
\end{eqnarray}%
has eigenvalues
$\bigg[4\alpha^2(1-\alpha^2)(e^{\omega_\mathbf{k}/T}+1)^{-1},0,0,0\bigg]$.
Thus, the concurrence is given by \begin{eqnarray}C(\varrho
_{A,II})=2\alpha\sqrt{1-\alpha^2}(e^{\omega_\mathbf{k}/T}+1)^{-\frac{1}{2}}
,\end{eqnarray} which is zero at zero Hawking temperature as
expected, and approaches the value $C_f(\varrho
_{A,II})=\alpha\sqrt{2(1-\alpha^2)}$ for infinite Hawking
temperature, which is just equal to $C_f(\varrho_{A,I})$ in this
limit. The entanglement of formation between mode $A$ and mode
$II$ is
\begin{eqnarray*}
E_{F}(\varrho_{A,II})=\mathcal
{H}\bigg[\frac{1+\sqrt{1-4\alpha^2(1-\alpha^2)(e^{\omega_\mathbf{k}/T}+1)^{-1}}}{2}\bigg]
\end{eqnarray*}
and the mutual information is
\begin{eqnarray}
I(\varrho_{A,II})&=&-\mathcal
{F}[1-\alpha^{2}(e^{-\omega/T}+1)^{-1}]-\mathcal
{F}[\alpha^{2}(e^{-\omega/T}+1)^{-1}]+\mathcal{F}(1-\alpha^{2})
\nonumber\\&&-\mathcal
{F}[1-\alpha^{2}(e^{\omega/T}+1)^{-1}]-\mathcal{F}[\alpha^{2}(e^{\omega/T}+1)^{-1}]+\mathcal
{F}(\alpha^{2}).
\end{eqnarray}
At $T=0$ the mutual information is zero, while in the infinite Hawking
temperature limit $T\rightarrow\infty$ the mutual information
becomes to $I_{f}(\varrho_{A,II})=-[\mathcal{F}(\alpha^{2})
+\mathcal{F}(1-\alpha^{2})]$, which is equal to
$I_{f}(\varrho_{A,I})$ in this limit.

We now study the entanglement between mode $I$ and mode $II$.
Tracing over the modes in $A$, we obtain the density matrix
\begin{eqnarray}
\nonumber\varrho_{I,II}=&&\alpha^2(e^{-\omega_\mathbf{k}/T}+1)^{-1}|00\rangle\langle00|+\alpha^2
(e^{\omega_\mathbf{k}/T}+e^{-\omega_\mathbf{k}/T}+2)^{-\frac{1}{2}}(|00\rangle\langle11|+|11\rangle\langle00|)\\
&&+(1-\alpha^2)|10\rangle\langle10|+\alpha^2(e^{\omega_\mathbf{k}/T}+1)^{-1}|11\rangle\langle11|,
\end{eqnarray}
where $|mn\rangle=|m\rangle_{I}|n\rangle_{II}$.

\begin{figure}[ht]
\includegraphics[scale=1.0]{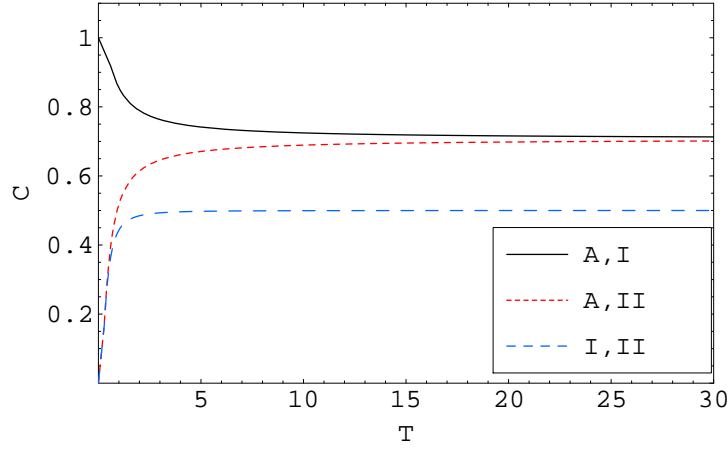}
\caption{\label{C}The concurrence $C(\varrho_{a,b})$ as a function
of the Hawking temperature $T$ with the fixed $\omega$ and
$\alpha$.}
\end{figure}

The partial transpose of $\varrho _{I,II}$ has an eigenvalue
\begin{eqnarray} \lambda_-
=-\frac{1}{2}[1-\alpha^2-\sqrt{1-2\alpha^2+\alpha^4+
\alpha^4(e^{\omega_\mathbf{k}/T}+e^{-\omega_\mathbf{k}/T}+2)
^{-1}}],\nonumber
\end{eqnarray} which is less than or equal to zero. Again, similar to
the last case, the entanglement does exist between these two modes
according to the partial transpose criterion. The matrix
$\varrho_{I,II}\,\tilde{\varrho}_{I,II}$ has eigenvalues
$[\alpha^4(e^{\omega_\mathbf{k}/T}+e^{-\omega_\mathbf{k}/T}+2)^{-1},0,0,0]$.
\ Thus, the concurrence is given by
\begin{eqnarray}C(\varrho_{I,II})=\alpha^2(e^{\omega_\mathbf{k}/T}+e^{-\omega_\mathbf{k}/T}+2)^{-\frac{1}{2}},
\end{eqnarray} which is zero at zero Hawking temperature, and
approaches the value $\alpha^2/{2}$ for infinite Hawking
temperature.

\begin{figure}[ht]
\includegraphics[scale=1.0]{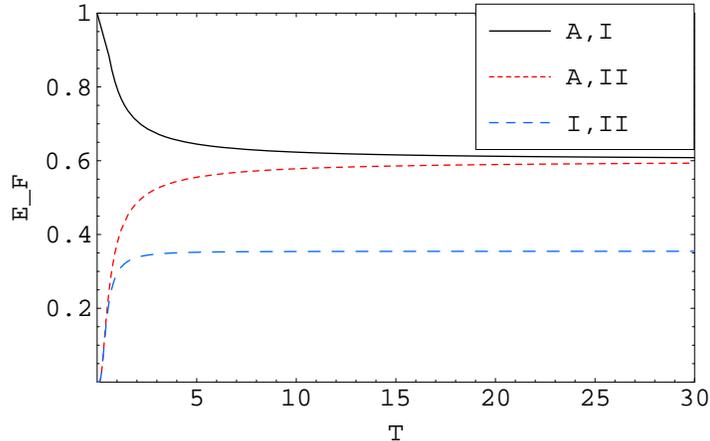}
\caption{\label{EF}The entanglement of formation $E(\varrho_{a,b})$
as a function of the Hawking temperature $T$ with the fixed $\omega$
and $\alpha$.}
\end{figure}

The entanglement of formation in this case is
\begin{widetext}
\begin{eqnarray}
E_{F}(\varrho_{I,II})=\mathcal
{H}\bigg[\frac{1+\sqrt{1-\alpha^4(e^{\omega_\mathbf{k}/T}+e^{-\omega_\mathbf{k}/T}+2)^{-1}}}{2}\bigg]
\end{eqnarray}
\end{widetext}
and the mutual information is%
\begin{eqnarray}
I(\varrho_{I,II})&=&\mathcal
{F}[1-\alpha^{2}(e^{\omega/T}+1)^{-1}]+\mathcal
{F}[\alpha^{2}(e^{\omega/T}+1)^{-1}]-\mathcal{F}(1-\alpha^{2})
\nonumber\\&&-\mathcal
{F}[1-\alpha^{2}(e^{-\omega/T}+1)^{-1}]-\mathcal{F}[\alpha^{2}(e^{-\omega/T}+1)^{-1}]-\mathcal
{F}(\alpha^{2}).
\end{eqnarray}
Again, we find that the mutual information  vanishes for zero
Hawking temperature, and increases to a finite value as the Hawking
temperature goes to infinity.

In Figs. (\ref{C}) and (\ref{EF})  we plot the behavior of the
concurrence and the entanglement of formation with the fixed
$\omega$ and $\alpha$ which show how the Hawking temperature would
change the properties of all the bipartite entanglement. When the
Hawking temperature is  lower, modes $A$ and $I$ remain almost
maximally entangled while there is little entanglement between modes
$I$ and $II$ and between modes $A$ and $II$. As the Hawking
temperature grows, the unaccessible entanglement between modes $I$
and $II$ and between modes $A$ and $II$ increases, while the
accessible entanglement between modes $A$ and $I$ degrades. We
arrive at the conclusion that the original entanglement in the state
Eq. (\ref{initial}) which is described  by the inertial observers is
now redistributed among the mode $A$ described by Alice, the mode $I$
in exterior region of the black hole described by Bob, and the
complimentary mode $II$ in the interior region of the black hole.
Therefore, as a consequence of the conservation of entanglement, the
physically accessible entanglement between the two modes described
by Alice and Bob is degraded.

\begin{figure}[ht]
\includegraphics[scale=1.0]{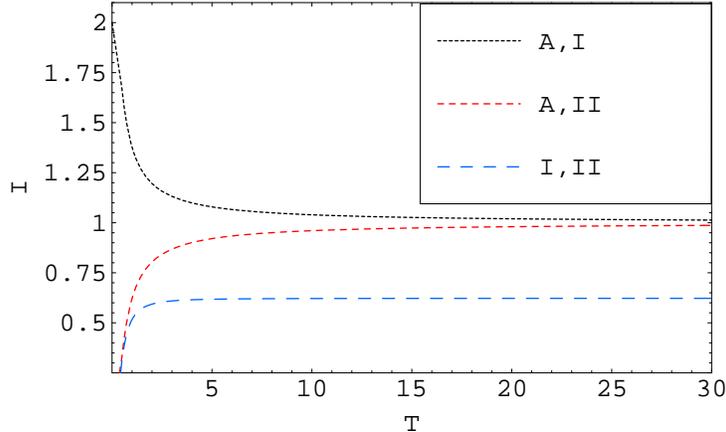}\vspace{0.0cm}
\caption{\label{MuIn} Mutual information $I(\varrho_{a,b})$ of the
Dirac modes versus Hawking temperature $T$ with the fixed $\omega$
and $\alpha$.}
\end{figure}

The properties of the mutual information are shown in Fig.
(\ref{MuIn}). It demonstrates that the mutual information of
$\varrho_{A,I}$ decreases while the mutual information of
$\varrho_{A,II}$ and $\varrho_{I,II}$ increases as the Hawking
temperature increases. It is interesting to note that when black
hole evaporates completely, the mutual information between modes $A$
and $I$ equals to just half of the its initial value and
$I(\varrho_{A,I})=I(\varrho_{A,II})$ in this limit, which are
independent of the state parameter $\alpha$.

In Ref. \cite{Pan Qiyuan} we found  that the entanglement and mutual
information of the mode $\varrho_{A,I}$ is degraded as the increase
of the Hawking temperature (or acceleration \cite{Schuller-Mann}).
The open question should be addressed in this paper is whether the lost correlations are destroyed or
transferred to somewhere. Here we presented a complete description
of the information behavior across an event horizon by discussing
the redistribution of the entanglement and mutual information. We
find that the correlations lost from the mode described by Alice and
the field mode outside the event horizon is gained by other
subsystems, especially between mode A and the mode $II$ inside the
event horizon. The results obtained here not only interpreted the
lose of entanglement and mutual information in the presence of a
horizon but also gave a better insight into the entanglement entropy
and information paradox of the black holes.

\section{summary}

The effect of Hawking radiation on the redistribution of entanglement 
in the Schwarzschild Spacetime is investigated. It is shown that the
entanglement between modes $I$ and $II$ and between modes $A$ and
$II$ increases, while the entanglement between modes $A$ and $I$ is
degraded  as the Hawking temperature grows. The original two-mode
entanglement, which is described by Alice and Bob from an inertial
perspective, is now redistributed among the mode $A$ described by
Alice, the mode $I$ in exterior region of the event horizon
described by Bob, and the complimentary mode $II$ in the interior
region of the horizon. This is a good explanation of physically
accessible entanglement between the two modes described by Alice and
Bob is degraded as the Hawking temperature grows. It is also found
that the mutual information of $\varrho_{A,I}$ decreases while
mutual information of $\varrho_{A,II}$ and $\varrho_{I,II}$
increases as the Hawking temperature increases.  It is interesting
to note that, in limit case that the temperature tends to infinity, the
accessible mutual information equals to just half of its initial
value, and the unaccessible mutual information between mode $A$ and
$II$ also equals to the same value. The results obtained here not
only interpreted the lose of entanglement and mutual information in
the presence of a horizon but also gave a better insight into the
entanglement entropy and information paradox of the black holes.

\begin{acknowledgments}

This work was supported by the National Natural Science Foundation
of China under Grant No 10875040; a key project of the National
Natural Science Foundation of China under Grant No 10935013; the
National Basic Research of China under Grant No. 2010CB833004, the
Hunan Provincial Natural Science Foundation of China under Grant No.
08JJ3010,  PCSIRT under Grant No IRT0964, and the Construct Program
of the National Key Discipline. Q. Y. Pan's work was partially
supported by the National Natural Science Foundation of China under
Grant No 10905020.

.
\end{acknowledgments}

\end{document}